\documentclass{iaus}
\usepackage{graphicx}

\title[Fast and Slow rotators] 
{Fast and Slow Rotators:\\ The build-up of the Red Sequence}

\author[Emsellem, Cappellari et al.]   
{Eric Emsellem,$^1$, Michele Cappellari,$^{2,3}$
  Davor Krajnovi{\'c},$^3$ 
  Glenn van de Ven,$^{2,4,5}$\thanks{Hubble Fellow}
  R.\ Bacon,$^1$ 
M.\ Bureau,$^3$ 
  Roger L.\ Davies,$^3$ 
  P.\ T.\ de Zeeuw,$^2$
  Jes\'us Falc\'on-Barroso,$^{2,6}$ 
 Harald Kuntschner,$^7$ 
  Richard McDermid,$^2$ 
  Reynier F.\ Peletier,$^{8}$
  Marc Sarzi,$^9$ 
  Remco C. E. van den Bosch,$^2$
}%
\affiliation{$^1$Universit\'e de Lyon~1,
Observatoire de Lyon; CNRS, UMR 5574, Centre de Recherche Astrophysique
de Lyon ; Ecole Normale Sup\'erieure de Lyon, Lyon,
France \\ [\affilskip]
$^2$Sterrewacht Leiden, Leiden University, Niels Bohrweg~2, 2333~CA Leiden, The Netherlands\\ [\affilskip]
$^3$Sub-Department of Astrophysics, University of Oxford, Denys Wilkinson Building, Keble Road, Oxford OX1 3RH, United Kingdom \\ [\affilskip]
$^4$Department of Astrophysical Sciences, Peyton Hall, Princeton, NJ 08544, USA \\  [\affilskip]
$^5$Institute for Advanced Study, Einstein Drive, Princeton, NJ 08540, USA \\ [\affilskip]
$^6$European Space and Technology Centre, Keplerlaan 1, Postbus 299, 2200 AG Noordwijk, The Netherlands \\
 [\affilskip]
$^7$Space Telescope European Coordinating Facility, European Southern Observatory, Karl-Schwarzschild-Str~2, 85748 Garching, Germany\\ [\affilskip]
$^8$Kapteyn Astronomical Institute, Postbus 800, 9700 AV Groningen, The Netherlands \\ [\affilskip]
$^9$Centre for Astrophysics Research, University of Hertfordshire, Hatfield, Herts AL10 9AB }

\pubyear{2007}
\volume{245}  
\pagerange{--}
\date{?? and in revised form ??}
\jname{Formation and Evolution of Galaxy Bulges}
\editors{M. Bureau, E. Athanassoula, and B. Barbuy, eds.}
\begin{document}

\maketitle

\begin{abstract}
Using the unique dataset obtained within the course of the {\tt SAURON} project, a
radically new view of the structure, dynamics and stellar populations of early-type 
galaxies has emerged. We show that galaxies come in two broad flavours (slow
and fast rotators), depending on whether or not they exhibit clear large-scale rotation, 
as indicated via a robust measure of the specific angular momentum of baryons.
This property is also linked with other physical characteristics of
early-type galaxies, such as: the presence of dynamically decoupled cores,
orbital structure and anisotropy, stellar populations and dark matter
content. I here report on the observed link between this baryonic
angular momentum and a mass sequence, and how this uniquely relates to the
building of the red sequence via dissipative/dissipationless mergers and
secular evolution. 
\end{abstract}

\firstsection 
\section{Red, Blue galaxies, Wet, Dry mergers}

The standard scenario for the formation of galaxy structures
includes hierarchical clustering of cold dark matter halos within which gas
is cooling (\cite[Peebles 1969]{Peebles69}; \cite[{Doroshkevich} 1970]{Dor70};\cite[{White} 1984]{White84}). 
The angular momentum of dark matter halos is thought to originate in cosmological torques and
major mergers (\cite[e.g., Vivitska et al. 2002]{Vivitska02+}).
If major mergers produce a significant increase in the 
specific angular momentum of the dark matter halos at large radii, 
minor mergers seem to just preserve or only slightly increase it with time
(\cite[Donghia et al. 2002, and Dongia et al., these proceedings]{Donghia02}). 
But little is known on the expected distribution of the {\em baryonic} angular 
momentum (\cite[{van den Bosch} et~al. 2002]{vdB02+}; 
\cite[{de Jong} et~al. 2004]{deJong04}), and even less if 
we focus on the central regions (within a few $R_e$).

The recent advent of large surveys such as the Sloan Digital Sky Survey allowed
to firmly establish a statistically significant bimodality in the colour
distribution of local galaxies, then separated in a so-called `blue cloud',
generally consisting of star-forming spiral galaxies, and a `red sequence',
mostly of non-star-forming early-type galaxies (e.g. \cite[Baldry et~al. 2006]{Baldry+04}). 
Accurately quantifying this bimodality (\cite[Bell et al. 2004]{}), 
allowed a dramatic improvement in the detailed testing of
galaxy formation scenarios. The bimodality can only be explained with the
existence of a feedback mechanism, which suppresses episodes of intense star
formation by evacuating the gas from the system. Many simulation groups have quantitatively
reproduced the bimodality, though with rather different assumptions for the
star formation and feedback processes (\cite[Springel et al. 2005]{Springel+05};
\cite[Cattaneo et al. 2006]{Cattaneo+06}; \cite[Bower et al. 2006]{Bower+06}).
A generic feature of these models is that red-sequence galaxies
form by dissipational `wet mergers' of gas-rich blue-cloud galaxies, followed
by quenching of the resulting intense star-formation caused by the feedback
from a central supermassive black hole and supernovae winds. The merging of
the most massive blue galaxies, however, is not sufficient to explain the
population of red-sequence galaxies, and dissipationless `dry mergers' of
gas-poor, red-sequence galaxies are also required, evolving galaxies along
the red-sequence as they increase in mass.

Wet and dry mergers both produce red, bulge-dominated galaxies, but the
kinematical structure of the remnants are expected to be very different. In a
merger between blue gas-rich galaxies, the gas tends to form a disk, so that the
end result of the merger, after the gas has been removed from the system
(ejection, conversion to stars), will be a red stellar system dominated by
rotation (e.g. \cite[Bournaud et al. 2005]{Bournaud+05}). In mergers between
red gas-poor galaxies, dissipationless processes dominate, resulting in a red
galaxy with little or no net rotation (e.g. \cite[Naab \& Burkert 2003]{Naab+03};
\cite[Cox et al. 2006]{Cox+06}). The existence of the red/blue galaxies
dichotomy therefore implies the existence of a kinematical differentiation
within the red sequence between fast and slow rotating galaxies. Various
observational indicators of this differentiation have been proposed in the past:
(i) anisotropy (from the $V/\sigma$ diagram, e.g. \cite[Davies et al. 1983]{Davies+83}), 
(ii) isophote shape (e.g. \cite[Kormendy \& Bender 1996]{Kormendy+96}), 
(iii) inner photometric slope (e.g. \cite[Lauer et al. 1995]{Lauer+95}).
However, none of these signatures
have been able to give clear evidence for a distinction
between the two classes of red-sequence galaxies,
primarily because they are secondary indicators of the
galaxies' internal dynamical structure.

\section{A new kinematic classification scheme}

By the application of integral-field spectroscopy to a representative sample of
nearby early-type galaxies, the {\tt SAURON} survey (\cite[Bacon et al. 2001]{Bacon+01};
\cite[de Zeeuw et al. 2002]{PaperII}) has revealed 
the full richness of the kinematics of these objects (\cite[Emsellem et al. 2004]{PaperIII}).
This unique dataset also allowed to robustly
distinguishing two distinct morphologies of stellar rotation fields,
corresponding to the predicted fast  and slow rotators
(\cite[Emsellem et al. 2007]{PaperIX}). We have thus defined a global
quantitative parameter, termed $\lambda_R \equiv \langle R \, \left| V \right| \rangle / 
\langle R \, \sqrt{V^2 + \sigma^2}\rangle $ linked to the baryonic
angular momentum, which shows that previous classification
schemes are not adequate (Paper~IX; \cite[Cappellari et al. 2007, Paper~X]{PaperX}). 

\begin{figure}
 \center{\includegraphics[width=6.7cm]{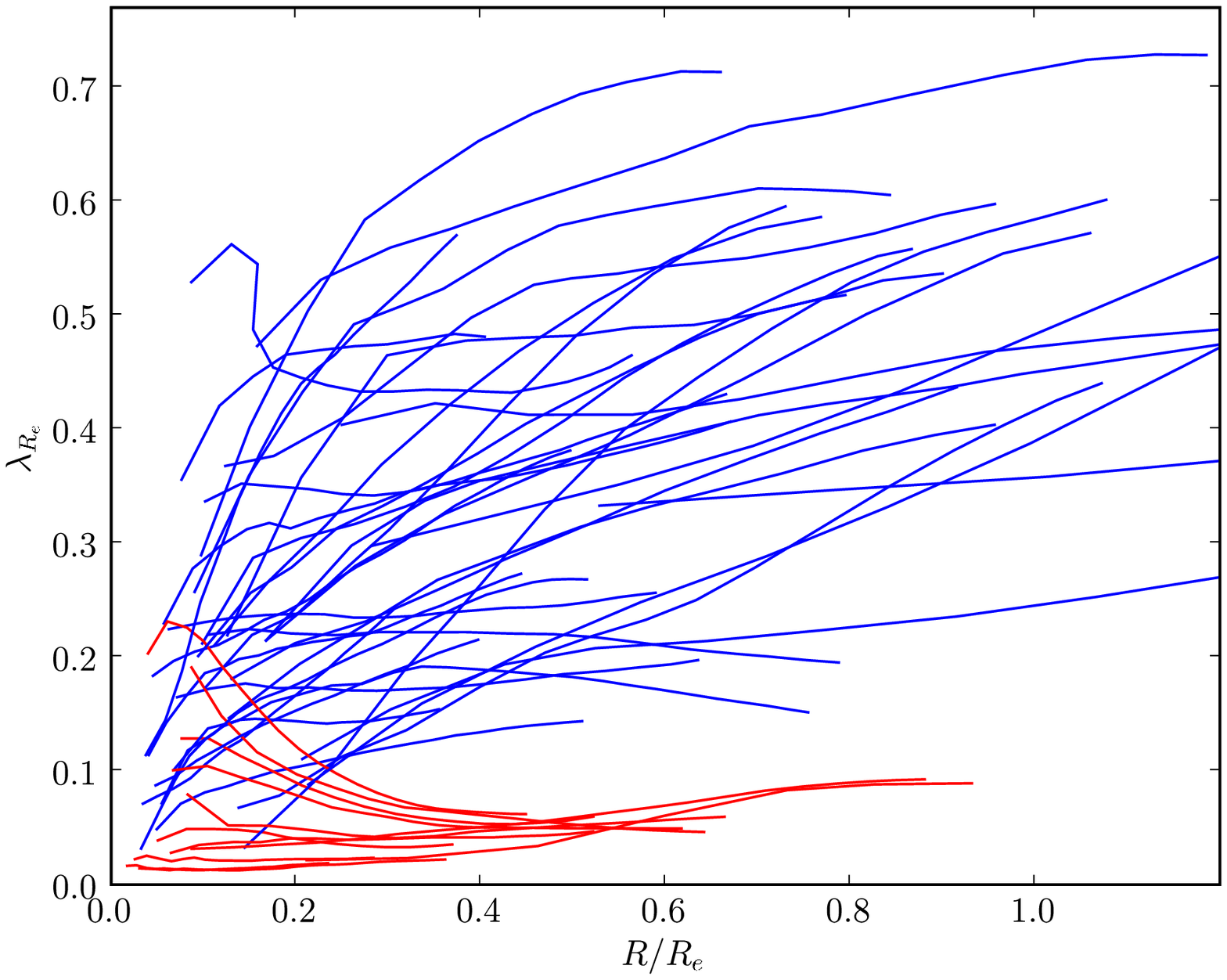}\includegraphics[width=6.7cm]{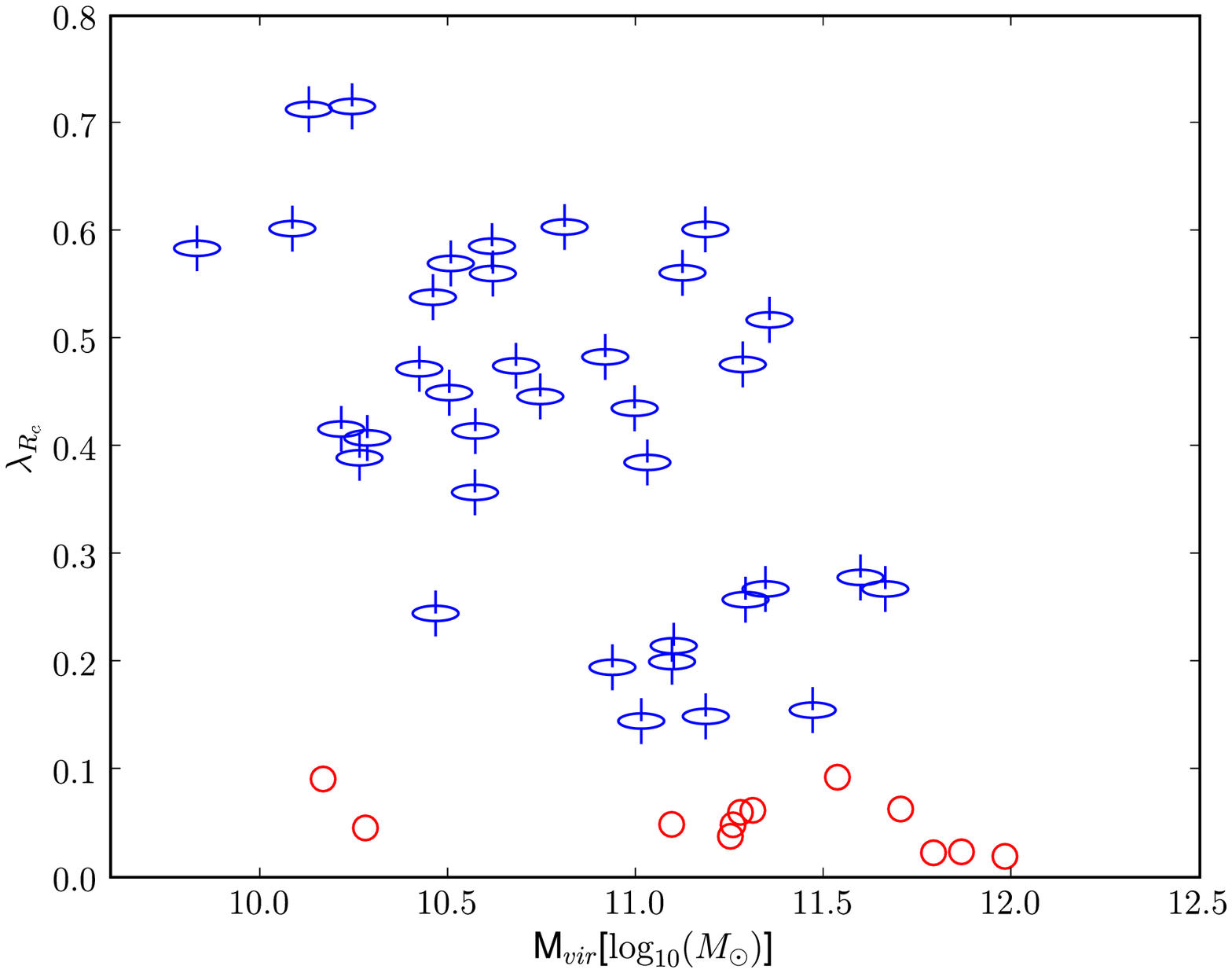}}
  \caption{Left panel: $\lambda_R$ profiles for our 48 E and S0 galaxies. Right panel: $\lambda_R$ at 1~$R_e$ versus
  the total mass of the galaxy. In both panels, slow and fast rotators are in red and
  blue, respectively.}\label{fig:lr}
\end{figure}

Fast and slow rotators are defined as having $\lambda_R$ values above or below 0.1, respectively.
Using the 48 E and S0 galaxies from the {\tt SAURON} sample, we can see a clear
difference between these two types of rotators: fast ones have rising $\lambda_R$
profiles, while slow rotators have either rather flat or decreasing profiles (Fig.~\ref{fig:lr}, left panel).
The apparent dichotomy may be partly due to the small number of available objects.
However, there are further indications that these two types of galaxies refer to 
different families. All fast rotators (except one galaxy with well-known irregular shells)
show well aligned photometric and kinemetric axes, 
and small velocity twists. This contrasts with most slow rotators which exhibit significant 
misalignments and velocity twists. These results are supported by a supplement of 
18 additional early-type galaxies observed with {\tt SAURON}. 

In Paper~X, we built state-of-the-art dynamical models of a subsample of 24 galaxies consistent
with axisymmetry, and found a trend with more intrinsically flattened galaxies 
being more anisotropic. The most massive galaxies are found {\em not} to be more anisotropic 
than the least massive ones. The results from these models are consistent with the 
distribution of all 48 galaxies of the {\tt SAURON} sample on the ($V/\sigma$,$\epsilon$) diagram. 
This was constructed for the first time from integral-field kinematics, 
using the revised and more robust formalism by Binney (2005). 
We showed that fast and slow rotators
have different distributions. Slow rotators are then more common among massive systems,
and are generally classified as E's from photometry alone. Fast rotators are
generally fainter and are classified either E or S0. 
These results strongly suggest that slow and fast rotators are really two different families
of galaxies. Fast rotators are consistent with being nearly oblate, and contain
disk-like components, while slow rotators are weakly triaxial. 

\section{The next steps}

Slow rotators, being on average more massive objects (Fig.~\ref{fig:lr}, right panel),  may result from 
mergers within the red sequence. The Kinematically Decoupled Components systematically
observed in slow rotators (except the ones which have $\lambda_R$ consistent with zero)
may well be the remnants of such a violent past history. However, current collisionless merger 
models seem unable to explain the detailed observed properties of
slow rotators (so far). It seems that a realistic merger tree may be a critical component
to obtain galaxies with such low baryonic angular momentum in their central
regions. The prediction is then that the angular momentum has been redistributed
among both the baryonic and dark matter components, but at larger radii, so
that we expect $\lambda_R$ profiles to shoot up outwards, even for slow rotators.

Fast rotators are more commonly obtained as the output of numerical simulations of
gas-rich mergers. Gas is a critical ingredient in the formation and evolution
of fast rotators, as illustrated by the presence of 100~pc-sized
decoupled (often counter-rotating) systems (\cite[McDermid et al. 2006]{PaperVIII}).
In this context, the E and S0 classification does not seem to be relevant, as
fast rotators form a continuous sequence of disky objects,
extending the bulge to disk ratios observed for spirals downwards.

$\lambda_R$ provides a robust classification of red-sequence galaxies that relates
directly to their formation history, and can be reproduced in cosmological
simulations. However, the galaxies in the original {\tt SAURON} survey were selected to
sample, with a relatively small number of objects, a wide
range of masses and shapes of early-type galaxies. Specifically, galaxies were
selected to be uniform in magnitude and ellipticity, but objects on the sky are
not uniformly distributed in these quantities. This selection imposed complex
biases and makes it impossible to derive a statistically meaningful distribution
of galaxy properties, for comparison with simulations. The latter requires
observing a statistically significant, volume-limited sample of galaxies
complete to some useful lower limit in mass. We therefore designed a new
ambitious two-dimensional spectroscopic survey, the {\tt ATLAS$^{\mathrm 3D}$},
of a magnitude limited sample of nearby early-type galaxies, resulting in
a complete sample of about 260 galaxies. This should
allow us to understand the distribution of Fast and Slow Rotators, to then infer
the relative fraction of wet / dry mergers, and further provide strong low-$z$ 
constraint on cosmologically motivated simulations.

\begin{acknowledgments}
I would like to thank the Conference organizers and IAU for financial support.
\end{acknowledgments}


\bibitem[\protect\citeauthoryear{{Vitvitska}, {Klypin}, {Kravtsov}, {Wechsler},
  {Primack} \& {Bullock}}{{Vitvitska} et~al.}{2002}]{Vivitska02+}
{Vitvitska} M.,  {Klypin} A.~A.,  {Kravtsov} A.~V.,  {Wechsler} R.~H.,
  {Primack} J.~R.,    {Bullock} J.~S.,  2002, \apj, 581, 799

\bibitem[\protect\citeauthoryear{{Peebles}}{{Peebles}}{1969}]{Peebles69}
{Peebles} P.~J.~E.,  1969, \apj, 155, 393

\bibitem[\protect\citeauthoryear{{Doroshkevich}}{{Doroshkevich}}{1970}]{Dor70}
{Doroshkevich} A.~G.,  1970, Astrophysics, 6, 320

\bibitem[\protect\citeauthoryear{{Rix} \& {White}}{{Rix} \&
  {White}}{1990}]{RixWhite90}
{Rix} H.-W.,  {White} S.~D.~M.,  1990, \apj, 362, 52

\bibitem[\protect\citeauthoryear{{van den Bosch}, {Abel}, {Croft}, {Hernquist}
  \& {White}}{{van den Bosch} et~al.}{2002}]{vdB02+}
{van den Bosch} F.~C.,  {Abel} T.,  {Croft} R.~A.~C.,  {Hernquist} L.,
  {White} S.~D.~M.,  2002, \apj, 576, 21

\bibitem[\protect\citeauthoryear{{de Jong}, {Kassin}, {Bell} \& {Courteau}}{{de
  Jong} et~al.}{2004}]{deJong04}
{de Jong} R.~S.,  {Kassin} S.,  {Bell} E.~F.,    {Courteau} S.,  2004, in
the proceedings of the International Astronomical Union Symposium no. 220, held 21 - 25 July, 
   2003 in Sydney, Australia. Eds: S. D. Ryder, D. J. Pisano, M. A. Walker, 
   and K. C. Freeman. San Francisco: Astronomical Society of the Pacific., p.281


\begin{thebibliography}{}
\bibitem[{Bacon} et~al. (2001)]{Bacon+01}
{Bacon} R.,  {Copin} Y.,  {Monnet} G., et al.  2001, \textit{MNRAS}, 326, 23 (Paper~I)

\bibitem[Baldry et al.(2004)]{Baldry+04} Baldry, I.~K., 
Glazebrook, K., Brinkmann, J., Ivezi{\'c}, {\v Z}., Lupton, R.~H., Nichol, 
R.~C., \& Szalay, A.~S.\ 2004, \textit{ApJ}, 600, 681 

\bibitem[Bell et al.(2004)]{Bell04} Bell, E.~F., et al.\ 2004, 
\textit{ApJ}, 608, 752 

\bibitem[Binney (2005)]{Binney05}
{Binney} J.  2005, \textit{MNRAS}, 363, 937

\bibitem[Bournaud et al.(2005)]{Bournaud+05} Bournaud, F., Jog, 
C.~J., \& Combes, F.\ 2005, \textit{A\&A}, 437, 69 

\bibitem[Bower et al.(2006)]{Bower+06} Bower, R.~G., Benson, 
A.~J., Malbon, R., Helly, J.~C., Frenk, C.~S., Baugh, C.~M., Cole, S., \& 
Lacey, C.~G.\ 2006, \textit{MNRAS}, 370, 645 

\bibitem[Cappellari et al.(2007)]{PaperX} Cappellari, M., et 
al.\ 2007, \textit{MNRAS}, 379, 418 (Paper~X)

\bibitem[Cattaneo et al.(2006)]{Cattaneo+06} Cattaneo, A., Dekel, 
A., Devriendt, J., Guiderdoni, B., \& Blaizot, J.\ 2006, \textit{MNRAS}, 370, 1651 

\bibitem[Cox et al.(2006)]{Cox+06} Cox, T.~J., Dutta, S.~N., 
Di Matteo, T., et al. 2006, \textit{ApJ}, 650, 791 

\bibitem[Davies et al.(1983)]{Davies+83} Davies, R.~L., 
Efstathiou, G., Fall, S.~M., Illingworth, G., \& Schechter, P.~L.\ 1983, 
\textit{ApJ}, 266, 41 

\bibitem[{de Jong} et~al. (2004)]{deJong04}
{de Jong} R.~S.,  {Kassin} S.,  {Bell} E.~F.,    {Courteau} S.  2004, in
proc. of the IAUS no. 220, Eds: S. D. Ryder, D. J. Pisano, M. A. Walker, 
   and K. C. Freeman. San Francisco: ASP, p.281

\bibitem[{de Zeeuw} et~al. (2002)]{PaperII}
{de Zeeuw} P.~T.,  {Bureau} M.,  {Emsellem} E.,  et al.  2002, \textit{MNRAS}, 329, 513 (Paper~II)

\bibitem[{D'Onghia} \& {Burkert}(2004)]{Donghia02}
{D'Onghia} E.,  {Burkert} A.  2004, \textit{ApJl} 612, L13

\bibitem[{Doroshkevich} (1970)]{Dor70}
{Doroshkevich} A.~G.,  1970, Astrophysics, 6, 320

\bibitem[{Emsellem} et~al. (2004)]{PaperIII}
{Emsellem} E.,  {Cappellari} M.,  {Peletier} R.~F.,  et al.  2004, \textit{MNRAS}, 352, 721 (Paper~III)

\bibitem[Emsellem et al.(2007)]{PaperIX} Emsellem, E., et al. 2007, \textit{MNRAS}, 379, 401  (Paper~IX)

\bibitem[Kormendy \& Bender(1996)]{Kormendy+96} Kormendy, J., \& 
Bender, R.\ 1996, \textit{ApJl}, 464, L119 

\bibitem[Lauer et al.(1995)]{Lauer+95} Lauer, T.~R., et al.\ 
1995, \textit{AJ}, 110, 2622 

\bibitem[{McDermid} et~al. (2006)]{PaperVIII}
{McDermid} R.~M.,  {Emsellem} E.,  {Shapiro} K.~L., et al.  2006, \textit{MNRAS}, 373, 906 (Paper~VIII)

\bibitem[Naab \& Burkert(2003)]{Naab+03} Naab, T., \& Burkert, 
A.\ 2003, \textit{ApJ}, 597, 893 

\bibitem[{Peebles} (1969)]{Peebles69}
{Peebles} P.~J.~E. 1969, \textit{ApJ} 155, 393

\bibitem[{Rix} \& {White} (1990)]{RixWhite90}
{Rix} H.-W.,  {White} S.~D.~M.  1990, \textit{ApJ} 362, 52

\bibitem[Springel et al.(2005)]{Springel+05} Springel, V., Di 
Matteo, T., \& Hernquist, L.\ 2005, \textit{MNRAS}, 361, 776 

\bibitem[{van den Bosch} et~al. (2002)]{vdB02+}
{van den Bosch} F.~C.,  {Abel} T.,  {Croft} R.~A.~C.,  {Hernquist} L.,
  {White} S.~D.~M.  2002, \textit{ApJ} 576, 21

\bibitem[{Vitvitska} et~al. (2002)]{Vivitska02+}
{Vitvitska} M.,  {Klypin} A.~A.,  {Kravtsov} A.~V.,  et al.  2002, \textit{ApJ} 581, 799

\bibitem[{White} (1984)]{White84}
{White} S.~D.~M.,  1984, \textit{ApJ}, 286, 38

\end{thebibliography}
\end{document}